\documentclass[aps,pra,superscriptaddress,nofootinbib]{revtex4}
\pdfoutput=1
\usepackage[intlimits]{amsmath}
\usepackage{amsfonts}
\usepackage{makecell}
\usepackage{enumitem}
\usepackage{psfrag}

\usepackage{amsthm}
\usepackage{float}
\usepackage{tikz}

\advance\voffset by -1cm

\usepackage[utf8]{inputenc}

\newcommand\beq            {\begin{equation}}
\newcommand\be            {\begin{equation}}
\newcommand\bea           {\begin{equation}\begin{array}l\displaystyle}
\newcommand\ee            {\end{equation}}
\newcommand\eeq            {\end{equation}}
\newcommand\bes           {\begin{subequations}}
\newcommand\esu           {\end{subequations}}

\renewcommand{\(}{\left(}
\renewcommand{\)}{\right)}
\renewcommand{\[}{\left[}
\renewcommand{\]}{\right]}

\counterwithout{table}{section}

\newcommand{\bigx}[1]{\bBigg@{#1}}

\def\3pt#1#2#3{{\langle{#1}\vert{#2}\vert{#3}\rangle}}

\newcommand\doi[2]        {\href{http://dx.doi.org/#1}{#2}}

\newcommand{\EQ}{\begin{equation}}
\newcommand{\EN}{\end{equation}}
\usepackage{epsfig}
\usepackage{psfrag}
\usepackage{amsmath}
\usepackage{graphicx}
\usepackage{amsfonts}
\usepackage{amssymb}

\def\tilde{\widetilde}
\def\bar{\overline}
\def\hat{\widehat}
\def\*{\star}
\def\[{\left[}
\def\]{\right]}
\def\({\left(}      

\def\){\right)}

\def\frac#1#2{\dfrac{#1}{#2}}

\def\2pi{\hbox{$2\pi i$}}

\def\dsl{\raise.15ex\hbox{/}\kern-.57em\partial}
\def\Dsl{\,\raise.15ex\hbox{/}\mkern-.13.5mu D}
\def\be{\beta}

\def\2pi{\hbox{$2\pi i$}}

\def\dsl{\raise.15ex\hbox{/}\kern-.57em\partial}
\def\Dsl{\,\raise.15ex\hbox{/}\mkern-.13.5mu D}

\def\barray{\begin{eqnarray}}
\def\earray{\end{eqnarray}}
\def\beq{\begin{equation}}
\def\eeq{\end{equation}}

\def\AA{\leavevmode\setbox0=\hbox{h}
\dimen0=\ht0 \advance\dimen0 by-1ex\rlap{\raise.67\dimen0\hbox{\char'27}}A}

\begin{document}

\title{Holographic Realization of the Prime Number Quantum Potential}

\author{Donatella Cassettari}
\affiliation{
SUPA School of Physics $\&$ Astronomy, University of St. Andrews, North Haugh, St. Andrews KY16 9SS, UK
}

\author{Giuseppe Mussardo}
\affiliation{SISSA and INFN, Sezione di Trieste, Via Bonomea 265, I-34136 
Trieste, Italy}

\author{Andrea Trombettoni}
\affiliation{Department of Physics, University of Trieste, Strada Costiera 11, I-34151 Trieste, Italy}
\affiliation{SISSA and INFN, Sezione di Trieste, Via Bonomea 265, I-34136
Trieste, Italy}

\begin{abstract}
  We report the first experimental realization of the prime number quantum potential $V_N(x)$, defined as the potential entering the single-particle Schr\"{o}dinger Hamiltonian with eigenvalues given by the first $N$ prime numbers. We use holographic optical traps and, in particular, a spatial light modulator to tailor the potential to the desired shape. As a further application, we also implement a potential with lucky numbers, a sequence of integers generated by a different sieve than the familiar Eratosthenes's sieve used for the primes. 
  Our results pave the way towards the realization of quantum potentials with arbitrary sequences of integers as energy levels and show, in perspective, the possibility to set up quantum systems for arithmetic manipulations 
  or mathematical tests involving prime numbers.  
\end{abstract}

\maketitle

\section{Introduction}
Mathematics is a golden mine of surprises, starting from its very basic branch: arithmetics. Consider as major examples the sets of the natural numbers 
\EQ
\mathbb{N}= \{1, 2, 3, 4, 5, \ldots\}
\EN 
and of the prime numbers 
\EQ
\mathbb{P}= \{2, 3, 5, 7, 11, \ldots\}.
\EN
While the pattern of natural numbers is obvious, since no matter which one you pick, it is straightforward to determine what the next one is,
the answer is instead highly non-trivial for the set of prime numbers, whose intriguing and sometimes apparently erratic properties never ceased
to intrigue mathematicians, physicists, scientists, and curious people in general \cite{Hardy,Apostol,Tao,Ore,Ribenboim,Schroeder,Zagier,Granville,Rose}. 

The sequences of integers and prime numbers are 
mathematical objects: one could say that they are the mathematical objects \textit{ par excellence}, 
being at the roots of arithmetics and therefore at the roots of entire mathematics. However, a useful point of view - particularly relevant for computational purposes -- is 
to see them as quantities which emerge from physical operations performed in the physical world. The rationale is to have a physical system, an \textit{abacus}, 
on which one performs the desired operations acting on the elements of certain desired sequences of integers. Since ultimately all aspects of the world around us can be explained using quantum mechanics, we would like to have 
a \textit{quantum abacus} and it is natural to employ a Hamiltonian whose eigenvalues are the elements of the desired quantum abacus.

Adopting this point of view, discrete sequences of numbers may be seen as spectra of some quantum Hamiltonians. In this paper 
we focus on the one-dimensional quantum Hamiltonian of a single particle of mass $m$, expressed in the standard form
as \EQ
\hat{H} \,=\, \frac{p^2}{2 m}  + V(x) \,\,\,,
\label{hamiltonian1d}
\EN 
where $p$ is the momentum operator and $V(x)$ is taken to be a continuous function in a given interval ${\mathcal J}$ (which can be the entire real axis). If a potential $V_N(x)$ is such that the eigenvalues $E_n$ of the time-independent Schr\"odinger equation
\EQ
\hat{H} \psi_n\,=\, \left(-\frac{\hbar^2}{2 m} \frac{d^2}{d x^2} + V_N(x) \right) \psi_n \,=\, 
E_n \psi_n \,\,\,
\label{ti_hamiltonian1d}
\EN
are the first $N$ prime numbers, we call $V_N(x)$ a \textit{$N$ prime number quantum potential}. In saying that the eigenvalues of \eqref{ti_hamiltonian1d} are the prime numbers, or any other sequence of integers, we are actually referring to the eigenvalues $e_n$ of $\hat{H}$ defined in dimensionless units: in physical units, the eigenvalues $E_n$ are equal to their dimensionless counterpart $e_n$ multiplied by a constant having the dimension of an energy, which depends on $\hbar$, $m$, and a length characteristic of the potential $V(x)$ itself. To fix the notation, hereafter the ground state in \eqref{ti_hamiltonian1d} corresponds to $n=0$ (and therefore, in the case of the primes, $e_0 =2$). For the sake of precision, let's point out that, in addition to the discrete part of the spectrum featuring the desired first $N$ primes, the potential $V_N$ will also have a continuous spectrum for energies larger than the highest considered prime $p_N$. 

The experimental realization of $V_N(x)$ will provide the first ingredient for the implementation of a quantum abacus, in which arithmetic operations can be translated into physical operations on the quantum particle. In this regard, light sculpting techniques provide versatile tools to dynamically control and engineer optical potentials of any desired form. These techniques are based on devices such as spatial light modulators (SLMs), digital micromirror devices (DMDs), and scanning acousto-optic deflectors (AODs) \cite{Gauthier}. In particular SLMs underpin computer-generated holography, in which a spatial phase modulation is applied to the trapping light such that a desired intensity distribution is realized in the far field. Holographic optical traps provide a flexible tool to tailor the potential experienced by neutral atoms, and have been employed in experiments ranging from single atoms to Bose-Einstein condensates \cite{Amico}. Therefore holographic techniques are a natural tool to implement the prime number quantum potential.

The aim of this paper is to present the first experimental realization of such a potential. As discussed in more detail in the next sections, the problem presents a number of simple but interesting points which touch on key theoretical and experimental questions in quantum mechanics and, at the same time, have fascinating outputs in number theory. The paper is organised as follows: in Section \ref{primessect} we recall some basic features of the prime numbers, pointing out the main challenges one has to face for setting up a potential $V(x)$ which has them as spectrum. In Section \ref{holsect} we provide a brief insight into holographic techniques.  In Section \ref{SQM} we present the theoretical framework of Supersymmetric Quantum Mechanics \cite{SUSYKhare,QMprime1,QMprime2} 
which leads to the exact expression of $V_N(x)$, and we also discuss the pros and cons of this approach compared to a semi-classical determination of the prime number potential \cite{GMscattering}. In Section \ref{experimental} we present the experimental prime number potential suitable for atom trapping, and we assess the feasibility of a subsequent implementation with ultracold atoms. In Section \ref{othersec}, to demonstrate the flexibility of computer-generated holography, we discuss the realization of another quantum potential associated to a different discrete sequence of integers, the so-called {\em lucky numbers}, an interesting set of integers which may be regarded as ``cousins'' of the prime numbers, generated by a slightly different sieve than the familiar Eratosthenes's sieve which gives rise to the prime numbers \cite{Ulamluckynumbers}. Our conclusions can be found in Section \ref{conclusions}. 

\section{The intriguing prime numbers}\label{primessect}
 A fundamental theorem of arithmetic states that every natural number greater than $1$ is either a prime number or can be represented as  product of prime numbers. Hence, the prime numbers may be regarded as the atoms of arithmetic but, in contrast with the finitely many chemical elements, the number of primes is instead infinite, as shown by a classic argument by Euclid dated more than 2000 years ago. Besides this fundamental role in arithmetic, what makes the prime numbers intriguing is their bipolar personality, i.e. in the realm of mathematics they are the perfect Dr. Jackill and Mr. Hyde. Such a mental insanity emerges by looking at the short and large distance scales of these numbers. Indeed, at short scale, their appearance along the sequence of the integers is completely unpredictable but, on a large scale, their coarse graining properties, and in particular how many prime numbers there are below any real number $x$, is an aspect which can be controlled with a great precision. In other words, while there is no known simple function $f(n)$ which gives the $n$-th prime number $p_n$ (and the actual determination of prime numbers can only be done by means of the familiar Eratosthenes's sieve \cite{Ribenboim,Schroeder,Zagier,Granville,Rose}), thanks to the insights of many prominent mathematicians (in particular Riemann), we have instead perfect knowledge of the inverse function $\pi(x)$ which counts the number of primes below the real number $x$ \cite{Hardy,Apostol,Tao,Ore,Ribenboim,Schroeder,Zagier,Granville,Rose,theorem1,theorem2,theorem3,theorem4,Riemannorig,Edwards}. Such a function has a staircase behaviour (since it jumps by $1$ each time that $x$ crosses a prime) but becomes smoother and smoother for increasing values of $x$. Its first estimate was empirically obtained by Gauss and Legendre 
\EQ
\pi(x) \sim \frac{x}{\ln x} \,\,\,,
\label{Gauss1}
\EN 
and, even though this formula may be considered just a coarse approximation of $\pi(x)$, it is nevertheless able to capture the asymptotic behaviour of $\pi(E)$ -- a result which constitutes the content of the ``Prime Number Theorem'' \cite{theorem1,theorem2,theorem3,theorem4}  
\EQ
\lim_{x\to\infty} \frac{\pi(x) \ln x}{x} \,=\,1 \,\,\,.
\label{theorem}
\EN
A more precise version of this estimate is given by $\pi(x) \simeq {\makebox {li}}(x) \equiv \int_2^x \frac{dt}{\ln t}$, 
while a further refinement was provided by Riemann \cite{Riemannorig,Edwards} in terms of the series 
\EQ
\pi(x)\, \simeq \,R(x) \, = \, \sum_{n=0}^{\infty} \frac{\mu(n)}{n}
\,{\makebox {li}}\left(x^{1/n}\right)  \,\,\, ,
\label{riemann}
\EN 
with the Moebius numbers $\mu(m)$ defined by 
$$
\mu(n) \,=\,
\left\{
\begin{array}{cl}
1 & \mbox {if $ n =1 $} \\
0 & \mbox {if $n$ is divisible by a square of a prime} \\
(-1)^k & \mbox{otherwise}
\end{array}
\right.
$$
where $k$ is the number of prime divisors of the integer $n$. It is worth stressing that $R(x)$ is the smooth function which approximates $\pi(x)$ more efficiently and it is well known that to reproduce the actual staircase jumps of $\pi(x)$ one needs to employs the zeros of the Riemann zeta-function 
\cite{Zagier,Riemannorig,Edwards}.

Knowing $\pi(x)$ helps us to estimate the growth behaviour of the $n$-th prime number. Indeed, by setting $p_n \,=\, \pi^{-1}(n)$ and inverting at the lowest order the function $\pi(x)$ (for instance using Gauss formula (\ref{Gauss1})), we have the following scaling law for the $n$-th prime number 
\EQ
p_n \simeq n \,\log n \,\,\,.
\label{scalinglaw}
\EN
However, the true unpredictable nature of the primes becomes particularly evident if we focus our attention on their gaps: for every prime $p_n$, let $g(p_n)$ be the number of composite numbers between $p_n$ and the next prime $p_{n+1}$, so that 
\EQ
p_{n+1} \,=\, p_n + g(p_n) + 1 
\,\,\,.
\EN 
With this definition, $g(p_n)$ is the size of the gap between $p_n$ and $p_{n+1}$. Using the scaling law (\ref{scalinglaw}), we expect the average gap $\bar{g}(p_n)$ between $p_n$ and $p_{n+1}$ to go as $\bar{g}(p_n)\sim\log n$, but the interesting question is: how wide is the range of values of these gaps? There is an extensive literature on this topic, see for instance \cite{Cramer,gap1,gap2,gap3,gap4}, and hereafter we only underline some basic features which are important for our subsequent considerations.  

The minimum value of $g(p)$ is $1$ and is obtained for the twin primes, i.e. the pairs such as $(17, 19)$ or $(29, 31)$, etc. which differ by $2$. Presently it is not known whether or not there are infinitely many twin primes, although there are strong reason to believe that the number of twin primes is indeed infinite (see, for instance, the heuristic arguments presented in \cite{Schroeder}). On the other hand, it is quite easy to show that $g(p_n)$ can be arbitrarily large, so that 
\EQ
\lim_{n\rightarrow \infty} {\rm sup}\, g(p_n)
 \,=\, \infty
\label{supgap}
\EN
To prove such a result, consider an arbitrary integer $N>1$ and the associated sequence of integers 
$$
N!+2, N!+3, N!+4, N!+5, ..., N!+N
$$
These $(N-1)$ consecutive numbers are all composite and therefore, if $p$ is the largest prime less than $N!+2$ we have $g(p) > N-1$.  Since we can send $N \rightarrow \infty$, we arrive to the result (\ref{supgap}). In summary, the sequence of primes shows a pattern of the gaps which is not at all regular, for instance one does not have any clue where the smallest gaps may appear. 

These features make evident the irregular behaviour of the primes and lead to the conclusion that the quantum potential $V_N(x)$ that encodes them should be a rather peculiar function. Given that it has $N$ bound states with energies equal to the prime numbers and strong variations in the energy gap between consecutive levels, we expect $V_N(x)$ to display a rich structure of maxima and minima which depends on $N$. Hence the experimental technique to realize $V_N(x)$ must be sufficiently flexible in order to accurately reproduce this structure. As shown in the next Section, experimental techniques such as computer-generated holography start with sampling $V_N(x)$ over a number of points ("pixels"). With more pixels available, it is possible to increase the complexity of $V_N(x)$, hence the number of energy levels. 

\section{Holographic techniques}
\label{holsect}

Before discussing how to obtain the analytic expression of $V_N(x)$, we review the experimental techniques which permit  the optical realization of the prime number potentials. Our optical potentials are suitable for trapping ultracold atoms in a one-dimensional geometry via the optical dipole force. The optical dipole potential is proportional to the intensity of the light \cite{Grimm}, hence here we shape the intensity profile of an incoming laser beam using holographic techniques. In computer-generated holography, a liquid-crystal SLM spatially modulates the phase of the light. The phase pattern on the SLM acts as a generalised diffraction grating, so that in the far field we have Fraunhofer diffraction and an intensity pattern is formed, which can be used to implement $V_N(x)$. The SLM acts effectively as a computer-generated hologram, and the light field in the output plane is the Fourier transform of the light field in the SLM plane. The calculation of the appropriate phase modulation to give the required output field is a well-known inverse problem which, in general, requires numerical solutions. Here we use a conjugate gradient minimization technique which efficiently minimizes a specified cost function \cite{Harte, Bowman}. The cost function is defined to reflect the requirements of the chosen light field in the output plane. In addition to specifying the intensity profile of the field, which gives $V_N(x)$, we also constrain the phase of the light in the output plane. Namely, a uniform phase is programmed across the whole intensity profile. Controlling the phase this way leads to a well-maintained intensity profile as the light propagates out of the output plane. 

\begin{figure}[t]
\centering
\includegraphics[width=0.8\textwidth]{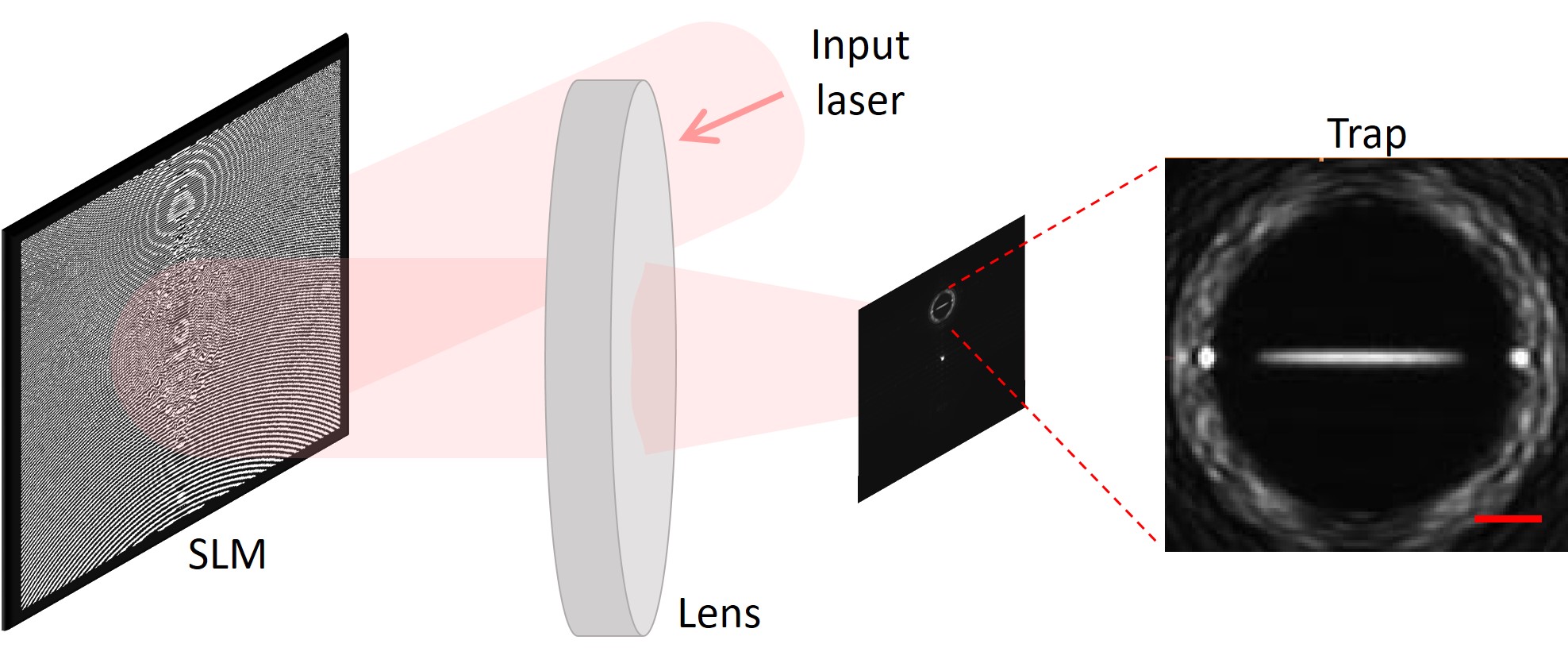}
\caption{Experimental setup, showing the phase profile imprinted by the SLM and the resulting light intensity profile in the trapping plane. The zoomed-in image shows the signal region, i.e. the region of the output plane in which the field is programmed by the conjugate gradient minimization algorithm. In this case the signal region contains a generic 1D trapping potential. The circular intensity distribution delimitates the boundary with the region of the output plane where the intensity is left unconstrained. The scale bar is 100 $\mu$m.} 
\label{ExpSetup}
\end{figure}

Figure \ref{ExpSetup} is a schematic of the experimental setup. Our SLM (Hamamatsu LCOS-SLM X10468) is illuminated by laser light with wavelength 1064 nm. The light diffracted by the SLM is focussed on the output plane by a $f=75$~mm achromatic doublet (Thorlabs AC508-075-B), and detected by a CCD camera (Thorlabs DCU224M). As shown in the figure, in order to accurately reproduce a given target light profile, we program only a small subset of the output plane (the "signal region" SR), whereas the field is left unconstrained in the rest of the plane
\cite{DeMarco}. We use the following cost function \cite{Bowman}:
\EQ
C = 10^{d}\left(1 - \sum_{p,q\in SR} \text{Re}\left\lbrace\left| \tilde{\tau}_{p,q}^{*} \tilde{E}^{\text{out}}_{p,q} \right|\right\rbrace\right)^{2}
\label{costfun}
\EN
where $p$ and $q$ denote the output plane coordinates. Here $\tau_{p,q}$ is the target electric field, $E^{\text{out}}_{p,q}$ is the output electric field, linked to the SLM electric field via a Fourier transform, and the over-tilde denotes normalization over the signal region. This cost function minimizes the discrepancy between $\tau_{p,q}$ and $E^{\text{out}}_{p,q}$ in the parameter space of all the different phase distributions that the SLM can generate. The prefactor $10^{d}$, where $d=9$ for the results shown here, increases the steepness of the cost function to improve convergence time and accuracy. 

In the algorithm, both the SLM plane and the output plane have a size of $2m\times 2m$ pixels, where $m\times m$ is the number of pixels in the SLM array: namely, the SLM plane is “padded” with zeroes to increase its size from $m\times m$ to $2m\times 2m$. The purpose of this is to fully resolve the output plane \cite{DeMarco}. We find that with $512\times 512$ SLM pixels available, we achieve 1D potentials that are up to 100-pixel long in the output plane. This amounts to about 1/10 of the linear dimension of the $1024\times 1024$ output plane. If we increase the length of the potential beyond this, we lose light-utilization efficiency, i.e.the light intensity in the pattern becomes too low. With our holographic method, we can obtain any smoothly varying intensity profile over this 100-pixel interval. However, given the non-trivial behaviour of the potential described at the end of the previous section, this 100-pixel maximum interval limits the number of energy levels of the potential, i.e. the number of primes. Therefore we envisage that to increase the number of primes contained in the potential beyond what we present in this work, it is necessary to increase the number of SLM pixels.

It will be possible to realize prime number potentials also with light sculpting techniques which are alternative to the liquid-crystal SLM we use in this work. For instance, AODs have been used to realize a wide range of optical potentials \cite{Trypogeorgos,Ryu}. While the AOD potentials realised so far are simpler than the prime number potentials presented here, it is possible in principle to increase their level of complexity. Another possibility is the use of DMDs. A DMD is a matrix of individually addressable mirrors which can be used as an intensity mask that can be directly imaged on the atoms. The projected image from a DMD is intrinsically binary, due to the individual mirrors being either “on” or “off”, however there are methods that overcome this limitation and allow the realization of intensity gradients: half-toning and time-averaging \cite{Gauthier}. Half-toning relies on the finite optical resolution of the imaging optical system, whereby multiple mirrors contribute to each resolution spot in the projected plane. This provides a number of possible intensity levels in each resolution spot which is given by the number of contributing mirrors. In \cite{Tajik} half-toning has been used to demonstrate 1D potentials with a high degree of control. If half-toning is combined with time-averaging, in which a time-averaged potential is achieved with high-speed modulation of the mirrors, intensity control can be further improved. With this combination of approaches, it will be possible to use DMDs to realize the prime number potentials presented here. The DMD dimension required for this is comparable to the dimension of the liquid-crystal SLM we use here, making the two techniques equivalent. 

\section{Discrete sequences and quantum potentials}\label{SQM}
Let's now address the problem of how to design, in general, a quantum potential $V(x)$ in such a way that a given sequence of real numbers   
\EQ
\mathcal{E} = \{e_0, e_1, e_2, e_3, \ldots\} 
\EN
coincides with the set of eigenvalues of the Schr\"{o}dinger equation (\ref{ti_hamiltonian1d}). It is important to distinguish two cases:
\begin{itemize}
\item the sequence $\mathcal E$ is {\em infinite}. In this case, a necessary condition for the existence of a continuous potential $V(x)$ able to support the sequence $\mathcal E$ as spectrum is that asymptotically the $e_n$'s satisfy the bound \cite{GMscattering}
 \EQ
e_n \,\leq \,A \,n^2 \,\,\,\,\,\,,\,\,\,\,\, n \rightarrow \infty
\label{creiterium}
\EN 
where $A$ is a positive real number. 
\item the sequence $\mathcal E$ is, on the contrary, {\em finite}, i.e. it only consists of a finite number $N$ of terms, $e_0, e_1, e_2, \ldots, e_{N-1}$, plus possibly a continuous part of the spectrum with energies larger than the maximum value of the $e_n$'s. 
In this case there is no obstacle to the existence of a potential $V(x)$ which supports such a spectrum, and the explicit form of this potential 
can indeed be found using methods of Supersymmetric Quantum Mechanics \cite{SUSYKhare}, as discussed below. 
\end{itemize}
A familiar example of the first case is provided by the infinite set $\mathbb{N}$ of all natural numbers, whose corresponding potential $V(x) \propto x^2$ gives rise to the well-known Hamiltonian of the harmonic oscillator \cite{Landau}. Another example is provided by the sequence $\mathcal{E} =\{1, 4, 9, 16, \ldots\}$ of squared integers, which can be realized as a quantum spectrum in terms of a properly tuned infinite-well potential \cite{Landau}. It is important to underline that, besides these known cases and very few others, it is in general {\em not} known a universal procedure for engineering a potential $V(x)$ with {\em exactly} all elements of an  infinite sequence $\mathcal{E} = \{e_0, e_1, e_2, e_3, \ldots\} $ as eigenvalues. The best one can do in such a case is to identify a {\em semiclassical} potential $V_{sc}(x)$: it is worth to stress, however, that this quantity is only able to capture the scaling growth of the eigenvalues rather their actual values since is determined by the formula 
\EQ
x(V_{sc}) \,=\,  \frac{\hbar}{\sqrt{2m}} \, \int_{E_{0}}^{V_{sc}} \frac{dE}{\frac{dE}{dn} \sqrt{V_{sc} -E}} \,\,\,,
\label{scpotential}
\EN
which depends upon the density of states $dE/dn$ rather than the individual energy levels $e_n$'s. Taking, for instance, the infinite set $\mathbb{P}$ of the primes, it is easy to see that these numbers  satisfy the bound (\ref{creiterium}) (see eq. (\ref{scalinglaw})). The corresponding semiclassical potential has been determined in \cite{GMscattering} by substituting in 
eq.\,(\ref{scpotential}) the density of states coming from (\ref{riemann}) 
\EQ
\left(\frac{dE}{dn}\right)^{-1} \,=\, \frac{d\pi}{dE} \simeq \frac{1}{\log E} \,\sum_{m=1}^{\infty} 
\frac{\mu(m)}{m} E^{(1-m)/m} \,\,\,.
\EN

Let's now consider the second case where the sequence $\mathcal{E} = \{e_0, e_1, e_2, e_3, \ldots\}$ is instead finite: here there exists the general procedure of Supersymmetric Quantum Mechanics (SQM) \cite{SUSYKhare} for engineering a potential ${\mathcal V}(x)$ with the $e_n$'s as its {\em exact} spectrum. It is well known that there are many potentials which share the same spectrum \cite{Kohn} and to identify uniquely one of these potentials, in the following we impose the additional condition ${\mathcal V}(x) = {\mathcal V}(-x)$. Using the  methods of SQM, one sets up a chain of potential ${\mathcal V}_k(x)$ ($k=N, N-1, \ldots 0$), as those shown in Figure \ref{chain}, with the property that the potential ${\mathcal V}_{k-1}(x)$ has the {\em same} spectrum of the previous one ${\mathcal V}_{k}(x)$ 
{\em except} its ground state. In other words, climbing down in the label $k$ of these potentials ${\mathcal V}_k(x)$, there is a depletion, one by one, of the lowest level of the previous potential. This chain of potentials is determined by a system of differential equations and, as we shall see below, this structure is at the root of the exact reconstruction of the potential with a given set of energy levels. Indeed, it is sufficient to reverse the procedure and adjust, one by one, all the desired eigenvalues! With a finite set of discrete eigenvalues, the final potential has a finite limit at $x\rightarrow \pm\infty$ and therefore also has a continuum part of the spectrum: however this is not relevant for our purposes and will not be discussed further. In more detail, the top-down procedure works as follows \cite{SUSYKhare}:

\begin{figure}[b]
\centering
\includegraphics[width=0.50\textwidth]{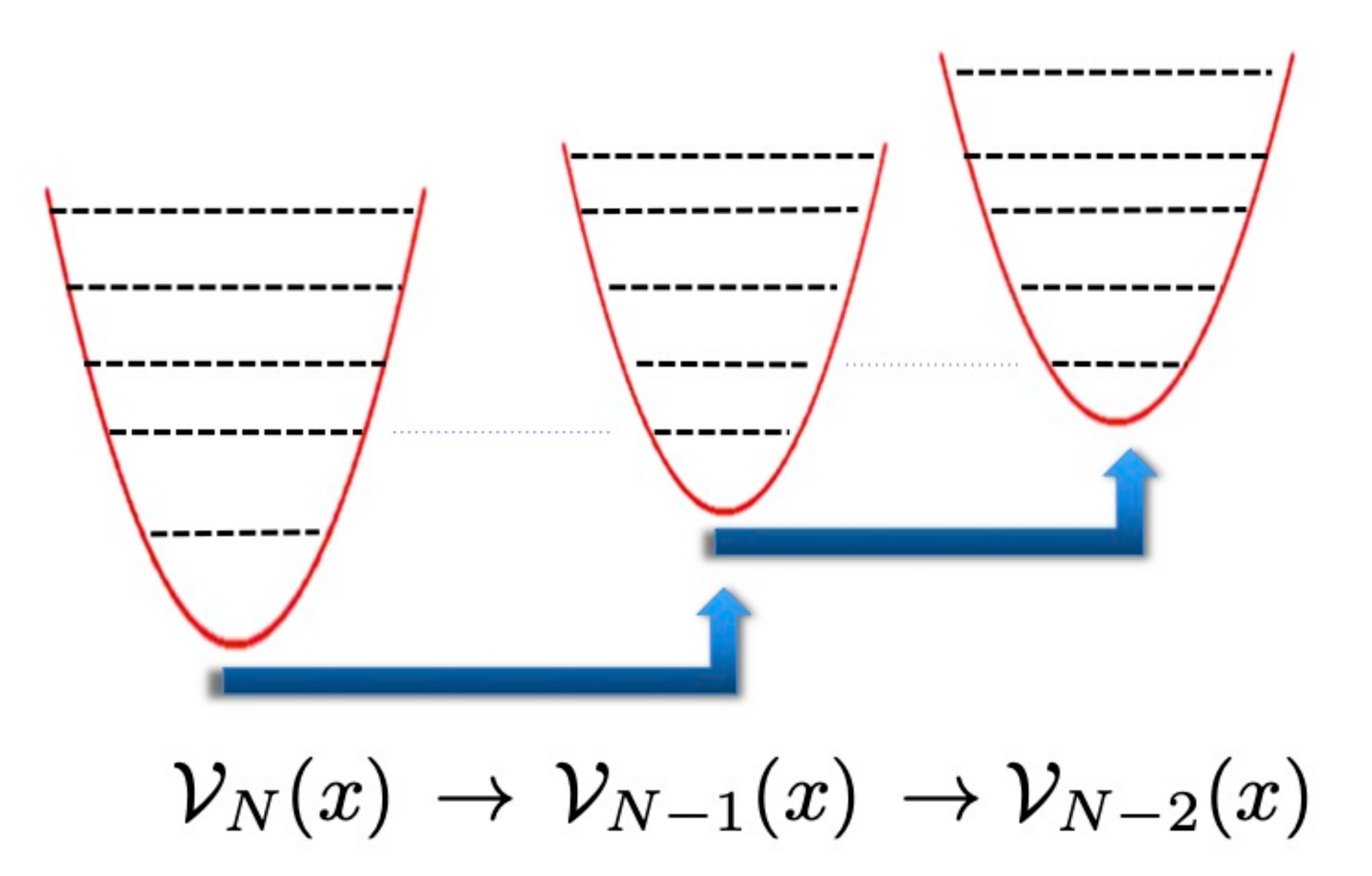}
\caption{Sequence of supersymmetric partner potentials ${\mathcal V}_k(x)$, where ${\mathcal V}_k$ shares all the spectrum of the previous one ${\mathcal V}_{k +1}$ except its ground state energy.} 
\label{chain}
\end{figure}

\begin{enumerate}
\item First of all, we subtract from all the eigenvalues $e_n$ the highest one $e_N$, so that the new set of numbers $\{\tilde{E}_n\}$ 
\beq
\tilde{E}_k \,=\, e_{N -k} - e_N 
\hspace{3mm} , \hspace{3mm} k=0,1,\ldots, N\,\,\,,
\label{newspectrum}
\eeq
will be considered as the new spectrum. The $\tilde{E}_k$'s are of course the (negative) gaps computed from the {\em highest} eigenvalue $e_N$. Notice that, consistently, they are enumerated starting from the top to the bottom, so ${\tilde E}_0=0$, ${\tilde E}_1$ is the first gap, ${\tilde E}_2$ the second gap, and so on. A potential ${\mathcal V}(x)$ where its only eigenvalue is 
${\tilde E}_0=0$ is of course ${\mathcal V}_0(x) =0$. This potential is used as input for the Riccati equation for the super-potential $W_1(x)$
\beq
W_1'(x) - W_1^2(x) + {\mathcal V}_0(x) \,=\, {\tilde E}_1 
\label{firsty}
 \eeq
with boundary condition $W_1(0)=0$.
\item Once such a function $W(x)$ has been obtained, one can construct another potential ${\mathcal V}_1(x)$ as 
\beq 
{\mathcal V}_1(x) \,=\, 2 {\tilde E}_1 + 2 W_1^2(x) - {\mathcal V}_0(x) \,\,\,.
\eeq
This potential is then substituted into (\ref{firsty}) (i.e., ${\mathcal V}_1(x) \rightarrow {\mathcal V}_0(x)$), substituting as well also 
${\tilde E}_1 \rightarrow {\tilde E}_2$, so that one has a differential equation for another super-potential $W_2(x)$ 
\beq
W_2'(x) - W_2^2(x) + {\mathcal V}_1(x) \,=\,{\tilde E}_2 \,\,\,,
\eeq
\item 
Proceeding iteratively in this way, one has a recursive sequence of differential equations 
\EQ
\begin{array}{l}
 W_k'(x) - W_k^2(x) + {\mathcal V}_k(x) \,=\, {\tilde E}_k  \,\,\,,\\
 \\
{\mathcal V}_k(x) = 2 {\tilde E}_{k} + 2 W_{k}^2(x) - {\mathcal V}_{k-1}(x) 
\end{array}
\label{diffeqs}
\EN
all of them solved with the boundary condition $W_k(0) = 0$, which ensures that the final potential ${\mathcal V}(x) = {\mathcal V}_N(x)$ is an even function. 
This recursive system is continued until all the gaps have been taken into account. Hence, solving (in general numerically) the differential equations (\ref{diffeqs}), one arrives to the Hamiltonian which has {\em exactly} the spectrum $\{e_n\}$ 
\beq
H\,=\, -\frac{d^2}{d x^2} + {\mathcal V}_N(x) + e_N \,\,\,,
\eeq
\end{enumerate}
These are indeed the theoretical steps 
which lead to the potential $V_N(x)$ with {\em exactly} the first $N$ prime numbers \cite{QMprime1,QMprime2}, in other words, our "quantum abacus'' with a finite number of beads. 
From what discussed above, the pros and cons of the SQM method are fairly evident: 
\begin{itemize}
\item {\em pros}: the method is {\em exact}, i.e. given a finite set of numbers $\{e_n\}$, this method provides the {\em exact} potential which has these numbers as the exact eigenvalues. If this set is made of the first $N$ primes, the potential exactly accommodates these prime numbers in the spectrum. 
\item {\em cons}: the method is {\em not} smoothly scalable, in the sense that if we have determined the potential which has as eigenvalues a set of $N$ values $e_n$, adding a new value $e_{N+1}$ we have to start again from scratch and determine altogether another potential which will accommodate {\em exactly} the $(N+1)$ eigenvalues. We will see below some examples of this feature. 
\end{itemize}
In comparison, the semi-classical potential $V_{sc}(x)$ (e.g. for the primes) has somehow opposite pros and cons: it is scalable, in the sense that once that it has been implemented, all its eigenvalues are fixed  but, on the other hand, it is not exact, i.e. its eigenvalues are not exactly the prime numbers. 
\begin{figure}[t]
\centering
\includegraphics[width=1.0\textwidth]{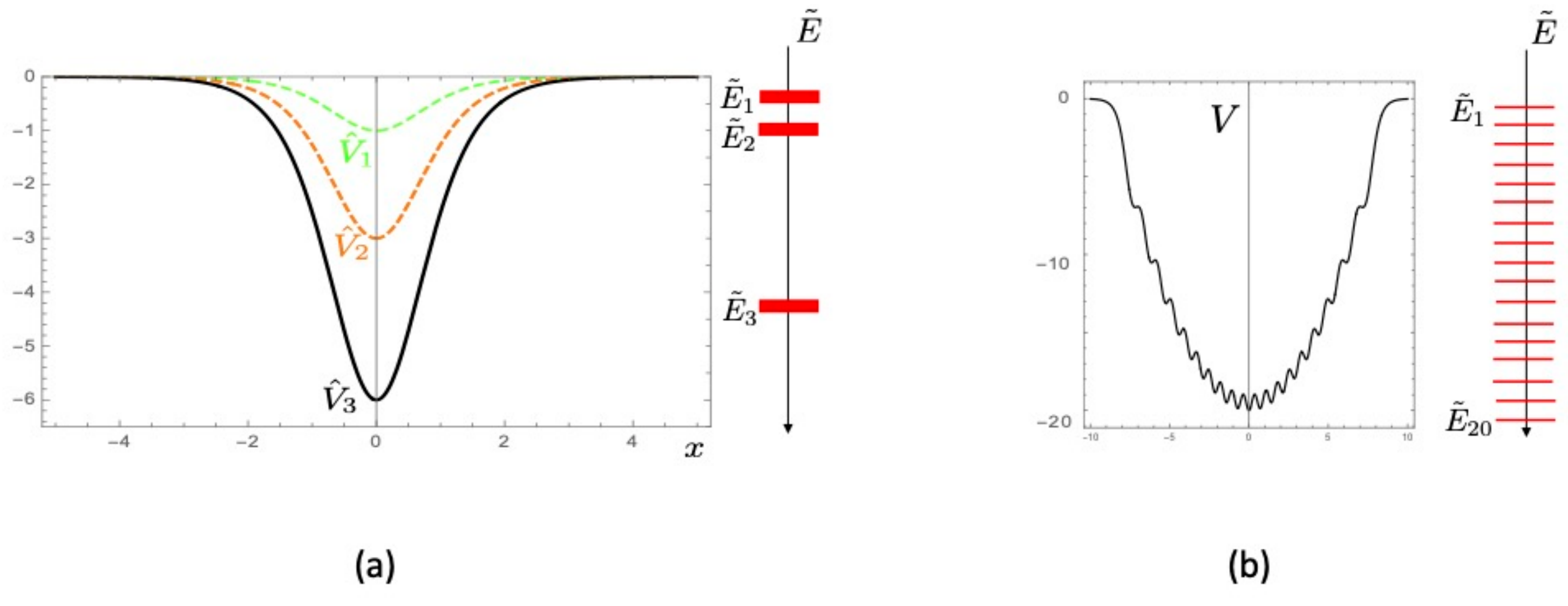}
\caption{(a) Sequence of P\"{o}schl-Teller potential $\hat V_N(x)$ obtained by solving the 
recursive differential equations (\ref{diffeqs}) to accommodate the energy gaps shown on the right hand side. (b) Potential $V(x)$ which has exactly the first (negative) $20$ natural numbers as energy gaps.} 
\label{PTpot}
\end{figure}

\begin{figure}[b]
\includegraphics[width=0.8\textwidth]{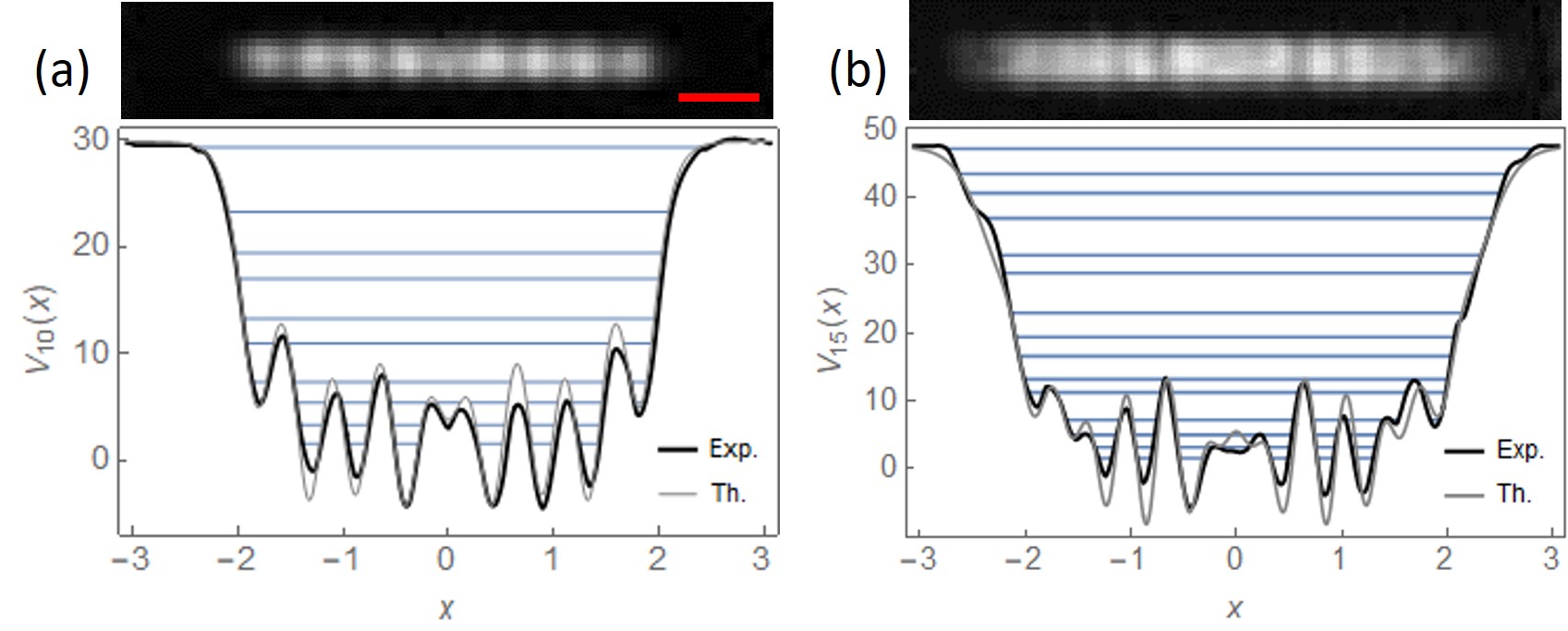}
\vspace{1mm}

 \centering
     \resizebox{0.8\textwidth}{!}{
    \begin{tabular}{ c|c|ccccccccccccccc| }
      \cline{2-17}    
\Large \textsf{(c)}\hspace{3mm} & Prime numbers & 2 & 3 & 5 & 7 & 11 & 13 & 17 & 19 & 23 & 29 & 31 & 37 & 41 & 43 & 47  \\ \cline{2-17}
& $e_{n}$ for $V_{10}(x)$ & 1.58 & 3.31 & 5.40 & 7.33 & 10.9 & 13.2 & 16.9 & 19.4 & 23.2 & 29.3 & & & & &  \\ \cline{2-17}
& $e_{n}$ for $V_{15}(x)$ & 1.58 & 3.21 & 5.00 & 7.22 & 11.3 & 13.2 & 16.6 & 19.4 & 22.9 & 28.8 & 31.4 & 36.9 & 40.6 & 43.4 & 47.1 \\ \cline{2-17}
    \end{tabular}}
\caption{(a,b) Experimental prime number potentials $V_{10}(x)$ and $V_{15}(x)$. The images show the light intensity profiles, where the red scale bar is 50 $\mu$m, applicable to both images. The plots show the corresponding potentials, re-scaled to dimensionless units, alongside their theoretical counterparts. The experimental eigenvalues are shown as horizontal lines on the plots and are also tabulated in (c), alongside the first 15 prime numbers for comparison.} 
\label{ExpRes}
\end{figure}

In closing this section, it is important to stress two important mathematical properties of the system of differential equations (\ref{diffeqs}). The first property is that there is one and {\em only one} family of potentials which provides a close analytic solutions of the system of the differential equations and which recursively reproduce themselves at each step of the procedure. This is the family of the P\"{o}schl-Teller potentials given by 
\EQ
\hat V_N (x)\,=\, - \frac{1}{2} \frac{N (N+1)}{\cosh^2 x}\,\,\,,
\EN
associated to the exact sequence of gaps 
\EQ
{\tilde E}_n = -\frac{n^2}{2} 
\,\,\,\,\,\,, \,\,\,\,\, 
n = 0, 1, 2, \ldots N
\label{PPTT}
\EN
The P\"{o}schl-Teller potentials $\hat V_N(x)$ do not have oscillations and present the typical shape of an inverted bell, see Figure \ref{PTpot}(a). The second property is that any finite sequence $\tilde{E}$ of gaps {\em other} than the one given by (\ref{PPTT}) is expected to give rise to a potential with {\em oscillations}. This is true even for very straightforward sequences, as for instance the potential which has exactly the sequence of the first $20$ negative integers as its energy gaps, see Figure \ref{PTpot}(b). This feature is also pretty evident in our realizations of the prime potentials: in Figure \ref{ExpRes} we show, for instance, the potentials $V_{10}(x)$ and $V_{15}(x)$ which have, as eigenvalues, exactly the first $10$ and $15$ prime numbers respectively: these potentials have a number of oscillations which scales as the number of accommodated eigenvalues. These oscillations are expected to be more pronounced in correspondence to more irregular sequences of numbers chosen to be energies. This is certainly the case for the prime numbers, as underlined in Section \ref{primessect}.


\section{Experimental prime number potentials}\label{experimental}

We use the apparatus shown in Figure \ref{ExpSetup} to realize the prime number potentials $V_{10}(x)$ and $V_{15}(x)$. The experimental light intensity profiles and the corresponding potentials are shown in Figure \ref{ExpRes}(a,b). The potentials are meant to be implemented with light that is red-detuned relative to the atomic transition, so that regions of higher intensity correspond to a lower value of the potential \cite{Grimm}. The potentials are re-scaled so that they are plotted in dimensionless units alongside their theoretical counterparts. The conversion between dimensionless eigenvalues $e_n$ and physical eigenvalues $E_n$ is given by:
\EQ
\frac{E_n}{e_n}=\frac{\hbar^2}{m}\left(\frac{l}{L}\right)^2
\EN
where $l$ and $L$ are the lengths of the potential in dimensionless and physical units respectively. Assuming $^{87}$Rb atoms, we obtain $E_n/e_n=h \times 0.029$~Hz for $V_{10}(x)$, and $E_n/e_n=h \times 0.026$~Hz for $V_{15}(x)$. The potential depth can be calculated with the same conversion formula, leading to a depth of $k_B\times 47$~pK for $V_{10}(x)$ and of $k_B\times69$~pK for $V_{15}(x)$. These figures show that we have shallow traps with very low trap frequencies, and that we would need to work at extremely cold temperatures, beyond what has been experimentally realized \cite{Medley, Leanhardt}. 

However trap depths and trap frequencies can be increased by increasing the numerical aperture of the optical system, i.e. with stronger focusing in conjunction with a larger SLM physical size. Our current optical system has a resolution spot of 10~$\mu$m. This determines the typical distance between peaks in the potentials, which is about three times the resolution spot. If in future we use a resolution of 1~$\mu$m, e.g. with a quantum gas microscope \cite{Haller,Zupancic}, then the distance between peaks and the overall physical length of the potentials can be scaled down accordingly. Given that energies scale with the square of the potential length, it will be possible to achieve trap depths of several nK, therefore improving experimental feasibility.

The experimental eigenvalues, expressed in dimensionless units, approximate the primes well: namely they match the primes if rounded to the nearest integer, as shown in Figure \ref{ExpRes}(c). The differences arise from the small discrepancies between the theoretical and experimental potentials, visible in Figure \ref{ExpRes}(a,b), which in turn are caused by an imperfect SLM response and by aberrations in the optical setup. The root-mean-square (r.m.s.) fractional discrepancy between the theoretical and the experimental potentials is 10\% for $V_{10}(x)$ and 7\% for $V_{15}(x)$, giving an r.m.s. fractional discrepancy between the eigenvalues and the primes of 8\% and 6\% respectively. In future these errors can be reduced with error-correction algorithms as shown in \cite{Gauthier}. Another source of experimental uncertainty is due to the optical power on the SLM fluctuating over time, as this will change the eigenvalues. Specifically, a $1\%$ fluctuation in the optical power leads to a $\sim 0.5\%$ change in the relative position of the eigenvalues. Hence to achieve the three-digit precision shown in Figure \ref{ExpRes}(c), it is necessary to stabilise the optical power to $1\%$, which is feasible with active stabilisation techniques.

\begin{figure}[t]
\begin{minipage}[b]{0.40\textwidth}
\includegraphics[width=0.98\textwidth]{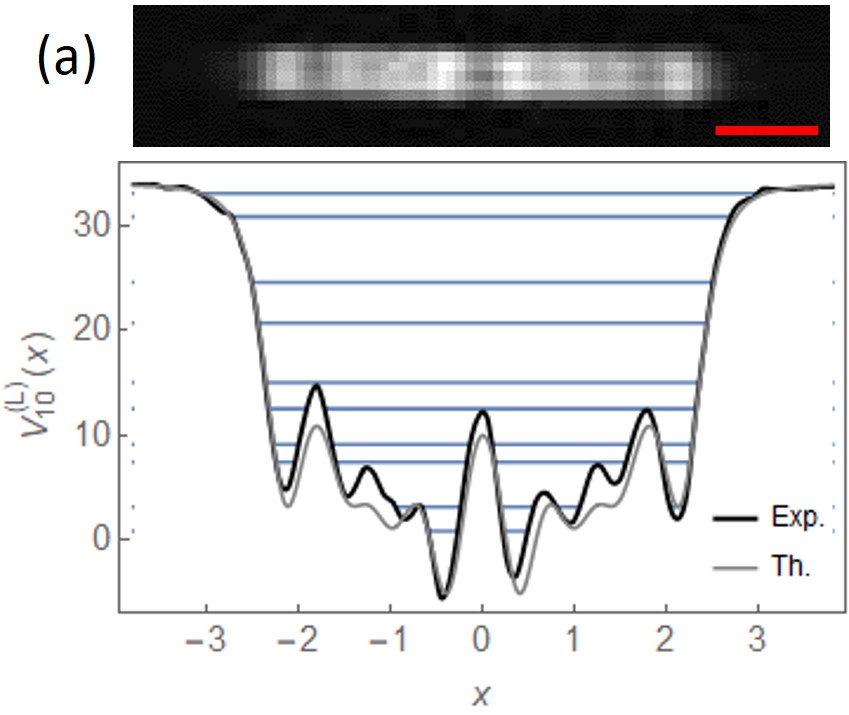}
\end{minipage}
\begin{minipage}[b]{0.22\textwidth}
\resizebox{\textwidth}{!}{
    \begin{tabular}{ c|c|c| }
      \cline{2-3}    
\large \textsf{(b)}\hspace{1mm} & \makecell{Lucky \\ numbers} & \makecell{$e_{n}$ for \\ $V_{10}^{(L)}(x)$} \\ 
\cline{2-3}
& 1 & 0.85 \\
& 3 & 3.16 \\
& 7 & 7.45 \\
& 9 & 9.12 \\
& 13 & 12.6 \\
& 15 & 15.1 \\
& 21 & 20.7 \\
& 25 & 24.6 \\
& 31 & 30.9 \\
& 33 & 33.2 \\
\cline{2-3} 
    \end{tabular}}
     \vspace{20pt}
\end{minipage}
\caption{(a) Experimental lucky number potential $V_{10}^{(L)}(x)$, where the red scale bar is 50 $\mu$m. The corresponding eigenvalues are tabulated in (b) alongside the first 15 lucky numbers.}
\label{potentiallucky}
\end{figure}
\section{The lucky quantum potential}
\label{othersec}
The method presented in this paper can be straightforwardly extended to other sequences of integers. As a significant example related to the primes, we present here the potential $V_N^{(L)}(x)$ having as eigenvalues the so-called lucky numbers 
\EQ
\mathbb{L}= \{1, 3, 7, 9, 13, 15, 21, 35, 31, 33, \ldots\} \,\,\, .
\EN
These numbers, introduced in the 50's by Gardiner, Lazarus, Metropolis and Ulam \cite{Ulamluckynumbers}, are obtained with a sieve (known as the sieve of Josephus Flavius) different from the sieve of Eratosthenes used for the primes. Briefly, to obtain the prime numbers  one notoriously eliminates from the list of integers the multiples of $2$
(the even numbers), then the multiples of $3$, then the multiples of $5$, and so on. On the contrary, for the lucky numbers one eliminates numbers based on their position in the remaining set, instead of their original value, i.e. their position in the initial set of natural numbers. So, one eliminates every second number (again the even numbers), then, rescaling the remaining set, every third number, then every fourth number, and so on. As for the primes, there are infinitely many lucky numbers. Moreover, the prime and the lucky numbers share many properties, including the asymptotic behaviour according to the prime number theorem. A "lucky prime" is a lucky number that is prime, and it has been conjectured that there are infinitely many lucky primes.

\begin{figure}[t]
  \centering
 \hspace{14mm} \includegraphics[width=0.8\textwidth]{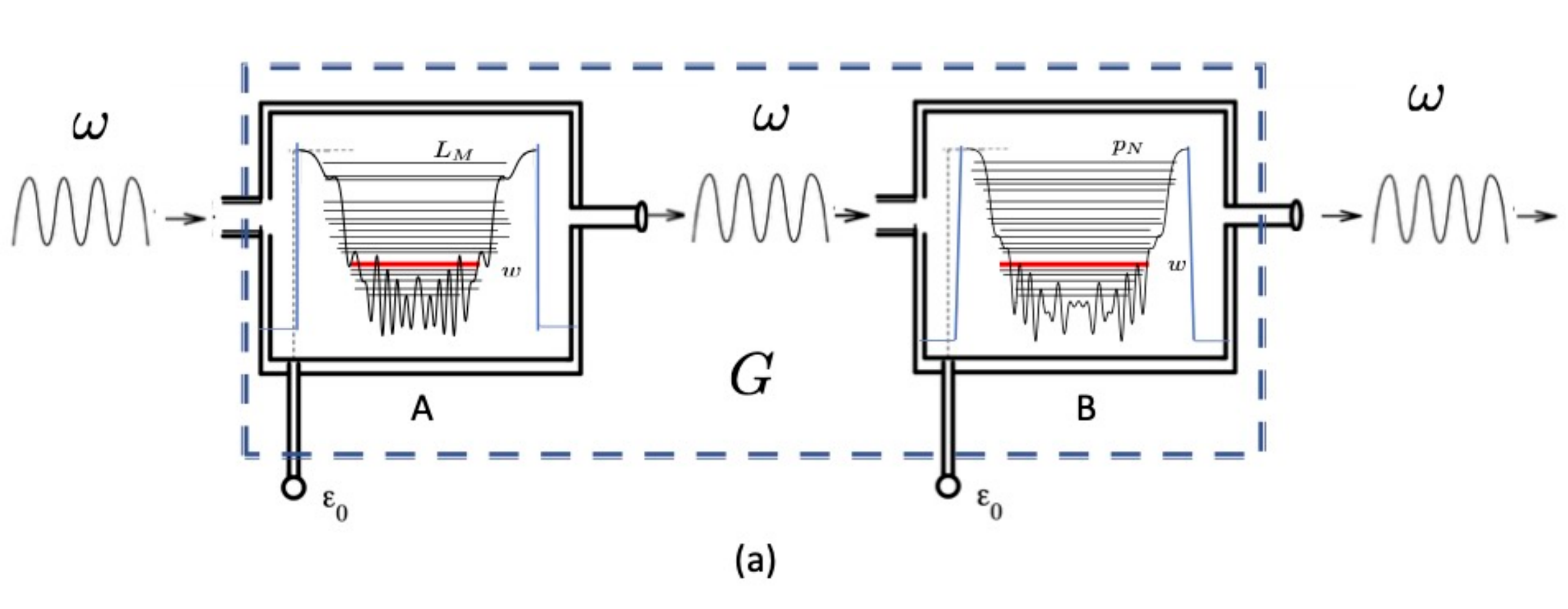}
 \centering
  \includegraphics[width=0.5\textwidth]{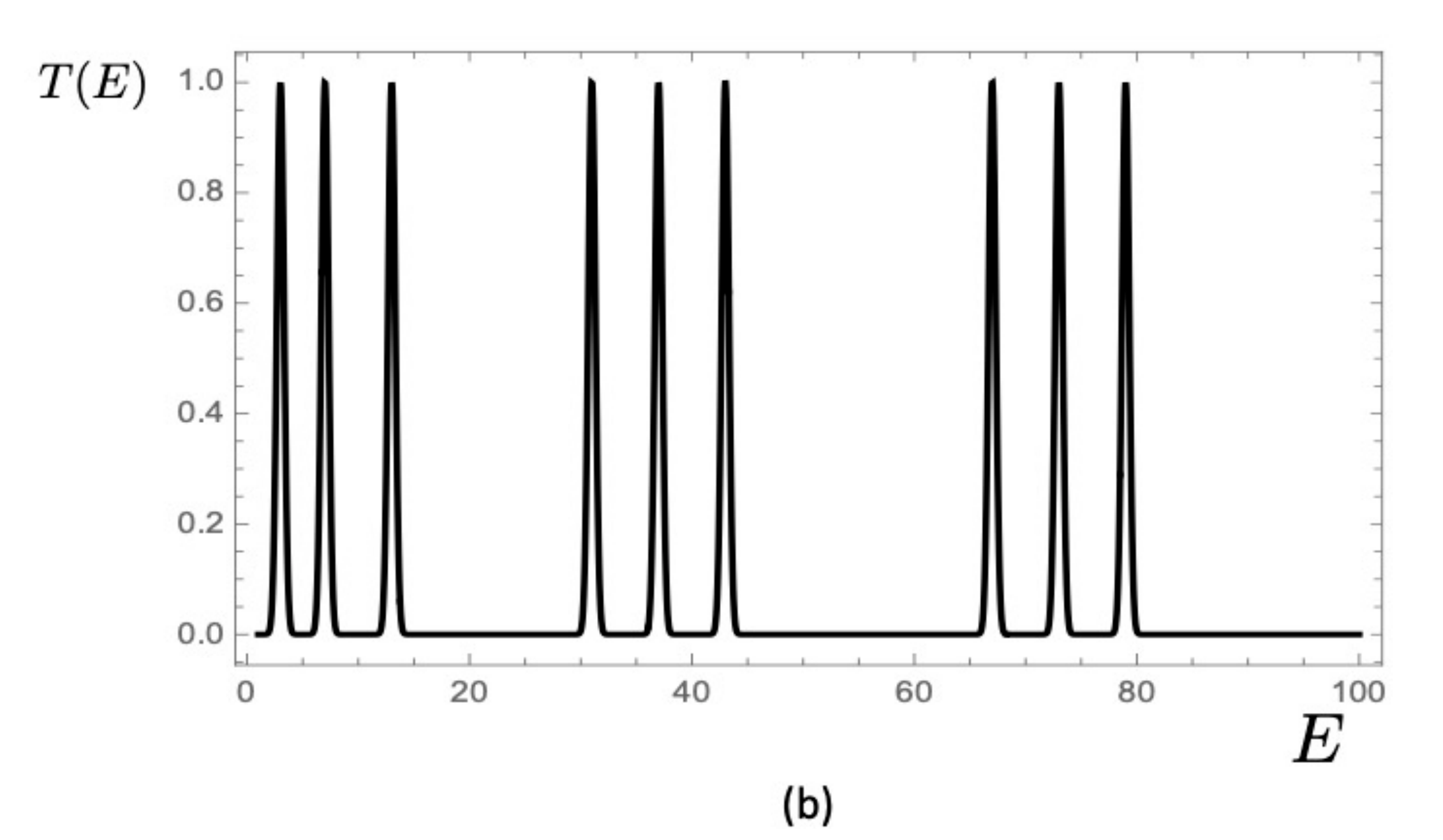}
   \caption{(a) The G-apparatus filters numbers that are both lucky and prime numbers. The devices A and B are made of the potentials $V_M^{(L)}(x)$ and $V_N(x)$ for the $M$ lucky numbers and $N$ prime numbers respectively, with an energy cutoff $\epsilon_0$. (b) The transmission amplitude $T(E)$ versus the (dimensionless) energy $E$ from the G-apparatus, with sharp resonance peaks in correspondence of those values of $E$ that are both lucky and prime numbers. }
\label{filterpotential}
\end{figure}

Proceeding as in Sections \ref{SQM} and \ref{experimental}, in Figure \ref{potentiallucky} we present, as an example, the experimental holographic realization of the potential $V_{10}^{(L)}(x)$ for the first $10$ lucky numbers. 
 Notice that, using the transmission and reflection properties of a quantum potential, it is possible to set up a simple physical experiment, shown in Figure \ref{filterpotential}, to test whether a given number ${\bf w}$ is both a lucky and a prime number. It involves a generalization of the proposal originally made in \cite{GMscattering} for checking the primality of a number: in the present case, let's imagine that in the box A we have realized the lucky potential $V_M^{(L)}$ with a number of levels $M$ large enough so that $L_M \gg {\bf w}$, while in the box B we have instead realized the prime number potential $V_N(x)$ with $p_N \gg {\bf w}$. Both potentials can be rounded and truncated at an energy cutoff $\epsilon_0$ (which can be controlled by an external handle) in such a way that the original energy levels are essentially left unperturbed but there are now asymptotic free states. Hence, we can take advantage of the typical resonance phenomena of quantum mechanics. We send on the composite apparatus $G$, made of A and B, a wave-packet from the left ($x \to -\infty$) with dimensionless energy ${\bf w}$. If the number ${\bf w}$ is a lucky number, it will be completely transmitted through box $A$, and if it is also a prime number, it will be completely transmitted through box $B$ as well. Therefore, if the particle with energy ${\bf w}$ is observed coming out the apparatus $G$, then the number ${\bf w}$ is both a lucky and a prime number. This way one could implement a experimental setup to test whether or not any given number ${\bf w}$ is a lucky prime.

\section{Conclusions}\label{conclusions}
In this paper we have provided the first experimental realization of
the prime number quantum potential $V_N(x)$, whose single-particle quantum Hamiltonian has the lowest $N$ prime numbers as eigenvalues. The exact theoretical shape of such a potential has been determined using supersymmetric quantum mechanics and experimentally implemented by means of holographic techniques. As a proof of principle, we have experimentally realized the potential $V_N(x)$ with $N=10$ and $N=15$, finding a  good agreement of the eigenvalues of these potentials with the first $15$ prime numbers. We have also discussed how this procedure can be successfully used to implement potentials having other sequences of integers as eigenvalues: this is the case of the ``lucky'' potential $V_N^{(L)}(x)$, i.e. the potential which has the first $N$ lucky numbers as eigenvalues.

  The present results provide a physical setup for a quantum mechanical manipulations of discrete sequences of numbers. This paves the way towards using these potentials for a variety of mathematical tests (such as the primality test) and arithmetic manipulations (such as prime factorization) by means of quantum experiments. It will be interesting to populate the energy levels with neutral atoms (bosonic or fermionic) and to induce transitions between levels by ``shaking'' the potential, either in terms of varying its overall strength or its center of mass, using a periodic drive. A compelling aspect is to determine whether it is better to employ for such manipulations either fermionic or bosonic atoms. Preliminary results seem to favour the latter, in absence of sizable atomic interactions, and further work is currently in progress. Equally interesting is to address other important open problems related to temperature effects and to the role played by atomic interactions, in view of the efficient implementation of any given arithmetic operation one wishes to implement on integers. 

\vspace{5mm}

{\textit Acknowledgements:} 
We would like to thank G. D. Bruce for technical assistance.

\end{document}